\documentclass[aps,prl, twocolumn,superscriptaddress]{revtex4-2}

\usepackage{xcolor}
\usepackage{graphicx}
\usepackage{dcolumn}
\usepackage{gensymb}
\usepackage{soul}

\begin{document}


\title{Fast and length-independent transport time supported by topological edge states in finite-size Su-Schrieffer-Heeger chains
}

\author{Yu-Han Chang}
\affiliation{Department of Physics, National Chung Hsing University, Taichung 402, Taiwan}

\author{Nadia Daniela Rivera Torres}
\affiliation{Department of Physics, National Tsing Hua University, Hsinchu 300, Taiwan}

\author{Santiago Figueroa Manrique}
\affiliation{Department of Physics, National Tsing Hua University, Hsinchu 300, Taiwan}

\author{Raul A. Robles Robles}
\affiliation{Institute of Photonics Technologies, National Tsing Hua University, Hsinchu 300, Taiwan}

\author{Vanna Chrismas Silalahi}
\affiliation{Department of Physics, National Chung Hsing University, Taichung 402, Taiwan}

\author{Cen-Shawn Wu}
\affiliation{Department of Physics, National Chang-Hua University of Education, Changhua 500, Taiwan}

\author{Gang Wang}
\affiliation{School of Physical Science and Technology, Soochow University, Suzhou 215006, China}

\author{Giulia Marcucci}
\affiliation{Department of Physics, University Sapienza, Rome 00185, Italy}
\affiliation{Institute for Complex Systems, National Research Council (ISC-CNR), Rome 00185, Italy}
\affiliation{Department of Physics, University of Ottawa, Ontario K1N 6N5, Canada}

\author{Laura Pilozzi}
\affiliation{Institute for Complex Systems, National Research Council (ISC-CNR), Rome 00185, Italy}
 \affiliation{Research Center Enrico Fermi, Via Panisperna 89a, 00184 Rome, Italy}

\author{Claudio Conti}
\affiliation{Department of Physics, University Sapienza, Rome 00185, Italy}
\affiliation{Institute for Complex Systems, National Research Council (ISC-CNR), Rome 00185, Italy}
\affiliation{Research Center Enrico Fermi, Via Panisperna 89a, 00184 Rome, Italy}

\author{Ray-Kuang Lee}
\email{rklee@ee.nthu.edu.tw}
\affiliation{Department of Physics, National Tsing Hua University, Hsinchu 300, Taiwan}
\affiliation{Institute of Photonics Technologies, National Tsing Hua University, Hsinchu 300, Taiwan}
\affiliation{Physics Division, National Center for Theoretical Sciences, Taipei 10617, Taiwan}
\affiliation{Center for Quantum Technology, Hsinchu 30013, Taiwan}

\author{Watson Kuo}
\email{wkuo@phys.nchu.edu.tw}
\affiliation{Department of Physics, National Chung Hsing University, Taichung 402, Taiwan}
\affiliation{Center for Quantum Technology, Hsinchu 30013, Taiwan}

\begin{abstract}
In order to transport information with topological protection, we explore experimentally the fast transport time using edge states in one-dimensional Su-Schrieffer-Heeger (SSH) chains. The transport time is investigated in both one- and two-dimensional models with topological non-trivial band structures. The fast transport is inherited with the wavefunction localization, giving a stronger effective coupling strength between the mode and the measurement leads. Also the transport time in one-dimension is independent of the system size. To verify the asertion, we implement a chain of split-ring resonators and their complementary ones with controllable hopping strengths. By performing the measurements on the group delay of non-trivially topological edge states with pulse excitations, the transport time between two edge states is directly observed with the chain length up to $20$. Along the route to harness topology to protect optical information, our  experimental demonstrations provide a crucial guideline for utilizing photonic topological devices. 
\end{abstract}
\maketitle

Composited by dimers with staggered hopping amplitudes,  a topological phase transition can be revealed in 
the Su-Schrieffer-Heeger (SSH) model owing to the existence of  Zak phase associated with zero Berry curvature
~\cite{SSH-1, SSH-2}.
By utilizing the SSH model, people have illustrated the difference between bulk and boundary, chiral symmetry, adiabatic equivalence, topological invariants, and bulk–boundary correspondence~\cite{book-SSH}.
Through the analogy in single-particle Hamiltonian, topologically non-trivial zero or $\pi$ modes can also be observed in photonic systems, through a periodically setting on the confining potentials~\cite{OL-Chen, TP-mat, TP-photon, TP-2D, Tao}.
With the topologically protected edge states, we can implement new types of lasing modes~\cite{laser1,laser} and optical control~\cite{control, nlin-SSH} even under continuous deformations.
As  the edge modes in a one-dimensional (1D) SSH model are zero-dimensional, by definition, they do not have a group velocity. 
However, for the practical implementation, instead of infinite chains, only a finite number of dimers can be fabricated. 
Moreover, in order to harness topology to protect optical information~\cite{tp-emitter}, a natural question arises on the corresponding  transport properties in the supported non-trivially topological states.
It is the common belief that the wavefunctions of supported edge states in a finite-size chain should remain staying localized strongly at their respective boundaries and decay exponentially in the bulk, with a penetration depth $\xi$ depending on the contrast between coupling strengths. On the contrary, in this paper, we reveal that one can transport information between two supported edge states with non-trivially topological protection with a longer distance, and achieving a much shorter transport time. Theoretically, the short transport time originates from the localization of edge modes, which do not scale with chain size, which is a major property of the transport time by the bulk modes. 

Experimentally,  the existence of non-trivially optical topological edge states in a chain of dimers with split-ring resonator (SRR) and its counterpart, the complementary split-ring resonator (CSRR), is illustrated with  a proper setting on the intra- and inter-cell coupling strengths between SRRs and CSRRs~\cite{Falcone04}. By extracting the amplitude and phase from the transmission  spectroscopy, a photonic band gap in the passband is measured  when the inversion symmetry is broken~\cite{Shelby01, Linden04}. The corresponding group velocity of edge states is directly measured with continuous wave and pulse excitations, verifying the transport of edge states when the system size is much larger than the penetration depth. Our results indicate that within a transport length the supported edge states possess not only non-trivial topologies, but also a scaling-free transport time.

The SSH model can be desrcibed by the Hamiltonian 
\begin{equation}
H_{SSH}=\sum_{j} \omega_0 a_j^\dag a_j+ \sum_{j} g_{j} a_{j+1}^\dag a_j+h.c.  
\end{equation}
where $a_j$ is the anhillation operator for site $j$, and $g_{j}=v$ and $w$ for odd and even $j$, respectively (Fig.1(a)). 
This staggered coupling in the ``diatomic basis" opens a band gap in the infinite chain case. The SSH model is known for its topological features and undergoes a transition from a trivial phase to a topological phase as it is tuned crossing $w/v  =1$. In the topological phase ($w/v > 1$), the model supports protected edge modes whose energy lies at the center of the band gap. Outside this band center, one recovers the typical Bloch solutions expected for periodic systems.
For a finite-size system a non-zero value of the energy difference $\Delta$ occurs for ``near-zero" modes even when $w/v > 1$. Given $w/v>1$, $\Delta$ can be approximated by 
$\Delta=2w (\frac{w}{v})^{-N/2}[1-(\frac{w}{v})^{-2}]/{[1-(\frac{w}{v})^{-N}]}$, in which $N$ is the chain length \cite{Efremidis}.
Associated with a finite $\Delta$, the corresponding modes are no longer isolated at one end, but become a symmetric or anti-symmetric combination of the isolated edge modes and characterized  by the penetration depth $\xi = 2/\ln(w/v)$ as presented in Fig. 1(b) ~\cite{Efremidis}. In this context, $\Delta/2$ can be viewed as a coupling energy of the localized modes on each edge,  scaled exponentially by $N$. 

To analyze the transport properties of the 1D chain, we model a setup where measurement leads are coupled to the ends of the chain. This configuration, combined with the gapped band structure of the SSH model, creates a system that resembles a typical molecular junction, with the chain itself acting as a long junction barrier. We can describe the photon transport to and from the leads by adding corresponding hopping terms to the Hamiltonian:
\begin{equation}
H=H_{SSH}+ g_l a_1+g_r a_N+h.c.,
\end{equation}
where $g_l$ is the hopping strength between the first resonator and the left lead, and $g_r$ is the hopping strength between the last resonator and the right lead.
For the transmission-type lead configuration shown in Fig. 1(a), the transmission coefficient $T(\omega)$ is related to the transmission susceptibility $\chi_T(\omega)$ by $T(\omega)=-i \sqrt{g_l g_r} \chi_T(\omega)$. The susceptibility is defined as \cite{Kockum13, Kuo19}:
\begin{equation}
\chi_T(\omega)=i\int_0^\infty dt e^{i\omega t} \left\langle \left[ a_N (t),\left( a_1+a_1^\dag \right) \right ]\right\rangle
\end{equation}
Here, $\left(a_1+a_1^\dag \right)$ represents the generalized force applied to the system at the first site, and $a_N (t)$ is the system's response observed at the last site. Similarly, the reflection susceptibility, $\chi_R(\omega)$, can be calculated by replacing the response operator $a_N(t)$ with $a_1(t)$ in Eq. (3).

The delay time for the photon transport via a resonance mode can be extracted from the phase data of $T(\omega)$ as detailed in Ref. \cite{Chang24}. By denoting $\kappa_c$ and $\kappa_i$ as the coupling decay rate and the intrinsic decay rate, respectively, and $\kappa = \kappa_c + \kappa_i$ the total decay rate, the maximum delay time associated with  transmission is $\tau_{d,m}={2}/\left({\kappa_i+\kappa_c}\right)={2}/{\kappa}$ at the resonance frequency $\omega=\omega_r$. Conversely, the reflected signal exhibits a negative delay, which indicates a pulse ``advance" with $\tau_{a,m}=-2{\kappa_c}/{\kappa_i(\kappa_i+\kappa_c)}$. Intriguingly, the difference between the maximum delay and the maximum advance depends only on the system's intrinsic loss rate $\tau_{d,m} - \tau_{a,m} ={2}/{\kappa_i}$. This result shows that achieving larger delay or advance times is contingent on using low-loss resonators.

In a system of coupled resonators, the projection of an eigenmode $|\Psi_s\rangle$ onto the contact sites plays a critical role in governing $T$ and $R$, by modifying the effective coupling decay rates for collective modes. For $R$, the effective coupling rate is $\kappa_c=\kappa_{c0}\left|\psi_1\right|^2$, in which $\psi_1=\langle 1|\Psi_s\rangle$ is the wavefunctions at lattice site $1$, and $\kappa_{c0}$ is a geometry-dependent coupling strength. For $T$, the total coupling decay rate involves both the input and output ports as $\kappa_c=\kappa_{in}+\kappa_{out}=\kappa_{c0}(\left|\psi_1\right|^2+\left|\psi_N\right|^2)$. Leveraging the parity symmetry of the modes ($\psi_N=e^{i\phi_s}\psi_1$), this can be approximated as $\kappa_c \sim 2\kappa_{c0}\left|\psi_1\right|^2$.

The effective $\kappa_c$ modifies the delay times for bulk modes to be $\tau_{d,m}={2}/\left({\kappa_i+\kappa_c}\right)={2}/\left({\kappa_i+2\kappa_{c0}|\psi_1|^2}\right).$ Close to the edge modes, there is the small splitting $\Delta$ comparable to $\kappa$ between them, and the maximum delay time is $\tau_{d,m}={2\gamma}/\left({\kappa_i+\kappa_{c}}\right)\sim {2\gamma}/\left({\kappa_i+\kappa_{c0}}\right),$ in which, $\gamma$ is a factor decreases from 2 to 1,  as $\Delta$ increases over $\kappa$. This doubling of the delay time is attributed to the superposition of the $2\pi$ phases contributed by each of the two hybridized edge modes. On the other hand, the maximum advance time for bulk modes is $\tau_{a,m}=-2{\kappa_{c0}|\psi_1|^2}/{\kappa_i(\kappa_i+\kappa_{c0}|\psi_1|^2)}.$ For edge modes $\tau_{a,m}\sim -{2}/{\kappa_i},$ by assuming $\kappa_{c0}>\kappa_i$, and the edge modes have significantly larger wavefunctions at the edge, with $\left|\psi_1\right|^2\sim 1$

In short, the key difference in delay times between edge and bulk modes stems from the wavefunction's amplitude at the contact site, $|\psi_1|$, and easy to understand in the ideal case $\kappa_i=0$. For bulk modes, we assume these behave like Bloch waves, with population equally distributed along the chain, $|\psi_1|^2 \sim 1/N$. The maximum delay time becomes $\tau_{d,m} = 1/(\kappa_{c0}|\psi_1|^2) \sim N/\kappa_{c0}$, scaling linearly with $N$. For edge modes, the wavefunction is localized at the boundary, giving $|\psi_1|^2 \sim 1$. This results in a maximum delay time of $\tau_{d,m} \sim 2\gamma/\kappa_{c0}$, which is $N$-independent. Athough the long delay time for bulk modes is a direct consequence of their small effective coupling rate, a classical interpretation is that the bulk mode transport is akin to a sequential, site-by-site hopping process, while edge mode transport resembles a single-step tunneling event. Meanwhile, the amplitude of the transmission, $T$, provides an estimate of the transfer efficiency for either process.

\begin{figure}[t]
\graphicspath{{fig/}}
\centering\includegraphics[width=0.47\textwidth, angle=0 ]{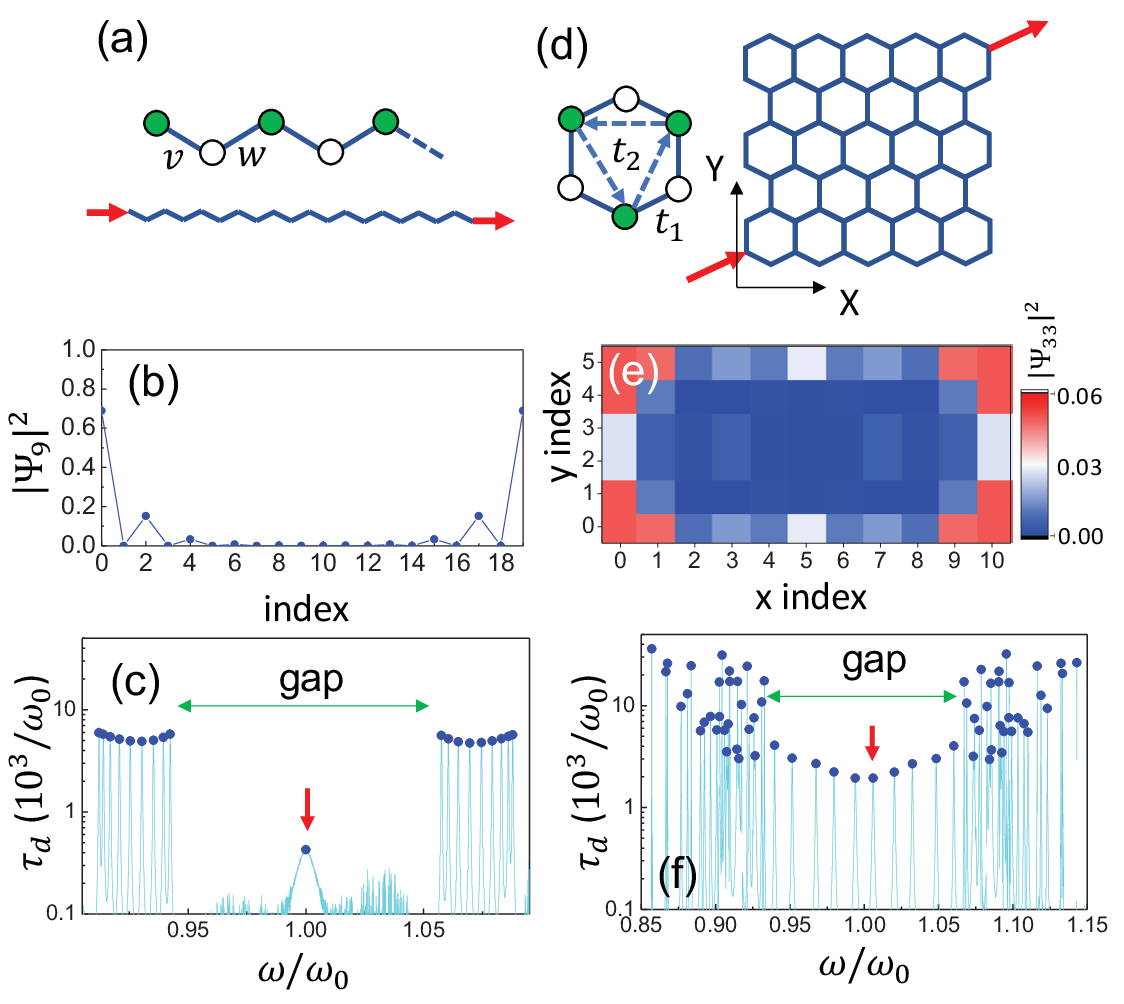}
\caption{(a) The SSH model with chain length $N=20$ and $v/\omega_0, w/\omega_0 = -0.01597, 0.07256$. (b) The corresponding edge modes $\Psi_9$ supported in this finite-sized SSH model. (c) Calculated delay time $\tau_d$ as a function of frequency with intrinisc loss $\kappa_i/\omega_0=10^{-4}$ and the coupling loss $\kappa_{c,0}/\omega_0=5\times 10^{-3}$. (d) The Haldane mode with $N= 11 \times 6$ and coupling strengths $t_1/\omega_0, t_2/\omega_0 = 0.05, 0.02j$. (e) The corresponding edge mode $\Psi_{33}$ supported in a finite-size Haldane mode. (f) Calculated $\tau_d$ as a function of frequency with $\kappa_i/\omega_0=10^{-5}$ and $\kappa_{c,0}/\omega_0=5\times 10^{-3}$. 
}
\end{figure}

Following Eq. (3), the transport delay $\tau_d = d\phi/d\omega$ is calculated and plotted in Fig 1(c), with blue symbols indicating $\tau_{d,m}$ at each resonance. The distribution of these resonances allows for the clear identification of the band gap and the topological edge mode, which is indicated by a red arrow. The central finding asserts that the edge mode transport has a much smaller delay time than the bulk mode transport. To confirm that this reduction in transport time is not restricted to the 1D case, we performed a similar calculation for a two-dimensional Haldane model (Fig. 1(d)) with $t_1/\omega_0, t_2/\omega_0 = 0.05, 0.02j$. The finite system consists of $5 \times 5$ honeycombs  and features a total of 66 modes. Figure 1(e) illustrates the wavefunction distribution ($|\Psi_{s}|^2$) for one of these edge modes $s=33$. The transport delay time for this model is shown in Fig 1(f). The distribution of resonances reveals the band gap and the 10 corresponding edge modes, in which the $s=33$ mode is marked by a red arrow. The result confirms again that the transport delay for the topological edge modes is significantly smaller than those of the bulk modes.

\begin{figure}[t]
\graphicspath{{fig/}}
\centering\includegraphics[width=0.47\textwidth, angle=0 ]{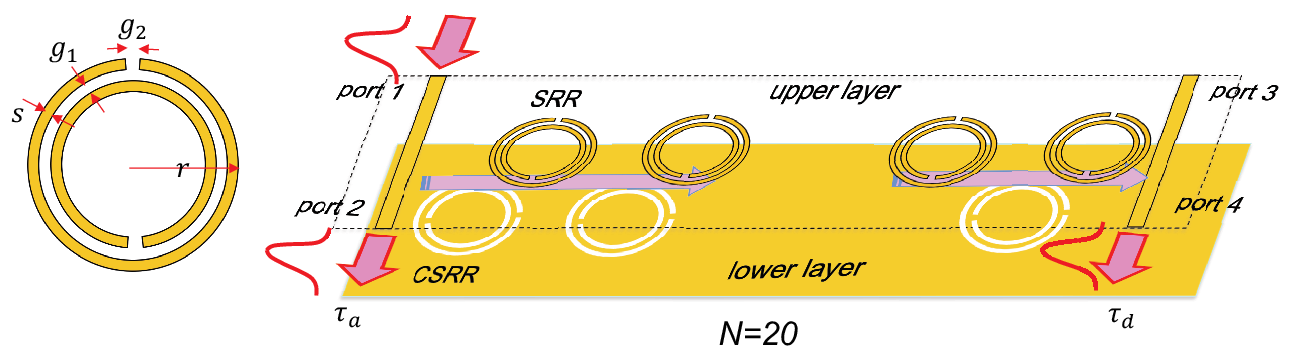}
\caption{
SSH model realized in a finite-size chain by combining SRRs and CSRRs. The two leads in the front and end nodes are implemented by two transmission lines, denoted with the input port $1$,  reflection port $2$, and transmissions ports $3$ and $4$, respectively.
}
\end{figure}

To demonstrate the fast transport supported by the topological edge states, we build 1D chains combining SRRs and CSRRs, fabricated on the front and back layers of a substrate, respectively, as illustrated in Fig. 2. Following the dimer structure, we manipulate the intra- and inter-cell coupling strengths by varying the orientation of SRRs and CSRRs. This control is possible because the coupling strength between SRRs is highly dependent on angle rotations \cite{Ser09, See17, Jiang}, and the coupling between an SRR and a CSRR is greatly impacted by their orientation texture \cite{Lin17, Chang23}. We fabricated two different types of samples: The superconducting Al resonators are fabricated on the substrate sapphire at a operation frequency of about 6 GHz, and the normal conducting Cu resonators are fabricated on the substrate Roger 4003C at a operation frequency of about 3 GHz. The Al resonators are designed with $1.406$ mm in diameter($r$), $0.148$ mm in line width($s$), gap between two rings($g_1$), and split for the ring gaps($g_2$), and $3.7$ mm for the lattice constant. The Cu resonators are designed with $r=7.6$ mm, $s=g_1=g_2=0.4$ mm, and $10$ mm for the lattice constant. To investigate the non-trivial topological case, we set the orientation angles of the SRR and CSRR to $(30\degree, 110\degree)$. For these selected angles, the resulting coupling ratio $w/v$ is $4.67$ for the Al resonators and $4.54$ for the Cu resonators. To perform direct measurements of the transmission and reflection spectra, we introduce two leads at the first and last nodes of the chain. With the help of the transmission lines, we define the input (port 1), reflection (port 2), and transmission (ports 3 and 4) ports.

\begin{figure}
\graphicspath{{fig/}}
\centering\includegraphics[width=0.4\textwidth, angle=0 ]{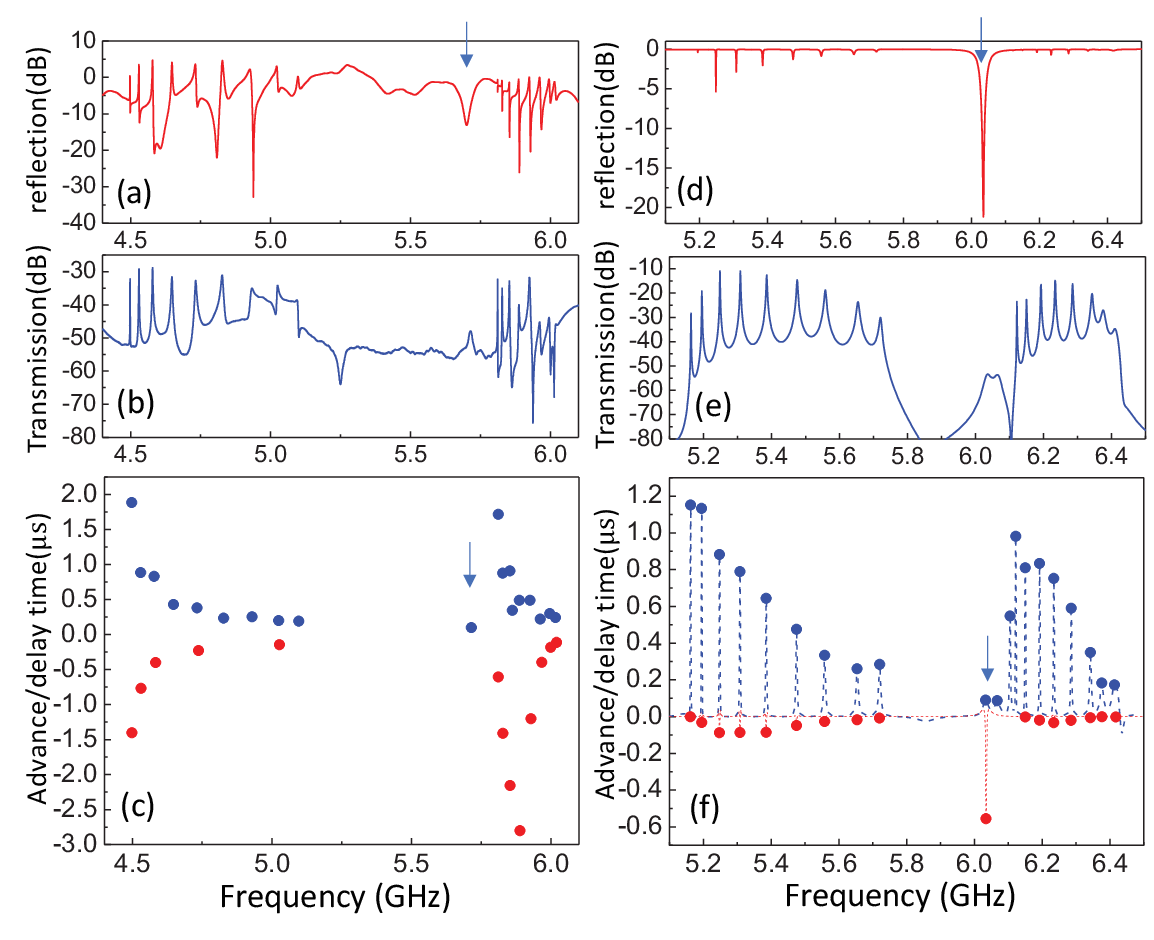}
\caption{(a) The measured reflection amplitude for a chain of 20 superconducting resonators. (b) The measured transmission amplitude. (c) The maximum advanced time(red) and delay times(blue) as a function of frequency. (d) The simulated reflection amplitude. (e) The simulated transmission amplitude. (f) The maximum advanced time(red) and delay times(blue) as a function of frequency. 
}
\end{figure}


\begin{figure}
\graphicspath{{fig/}}
\centering\includegraphics[width=0.4\textwidth, angle=0 ]{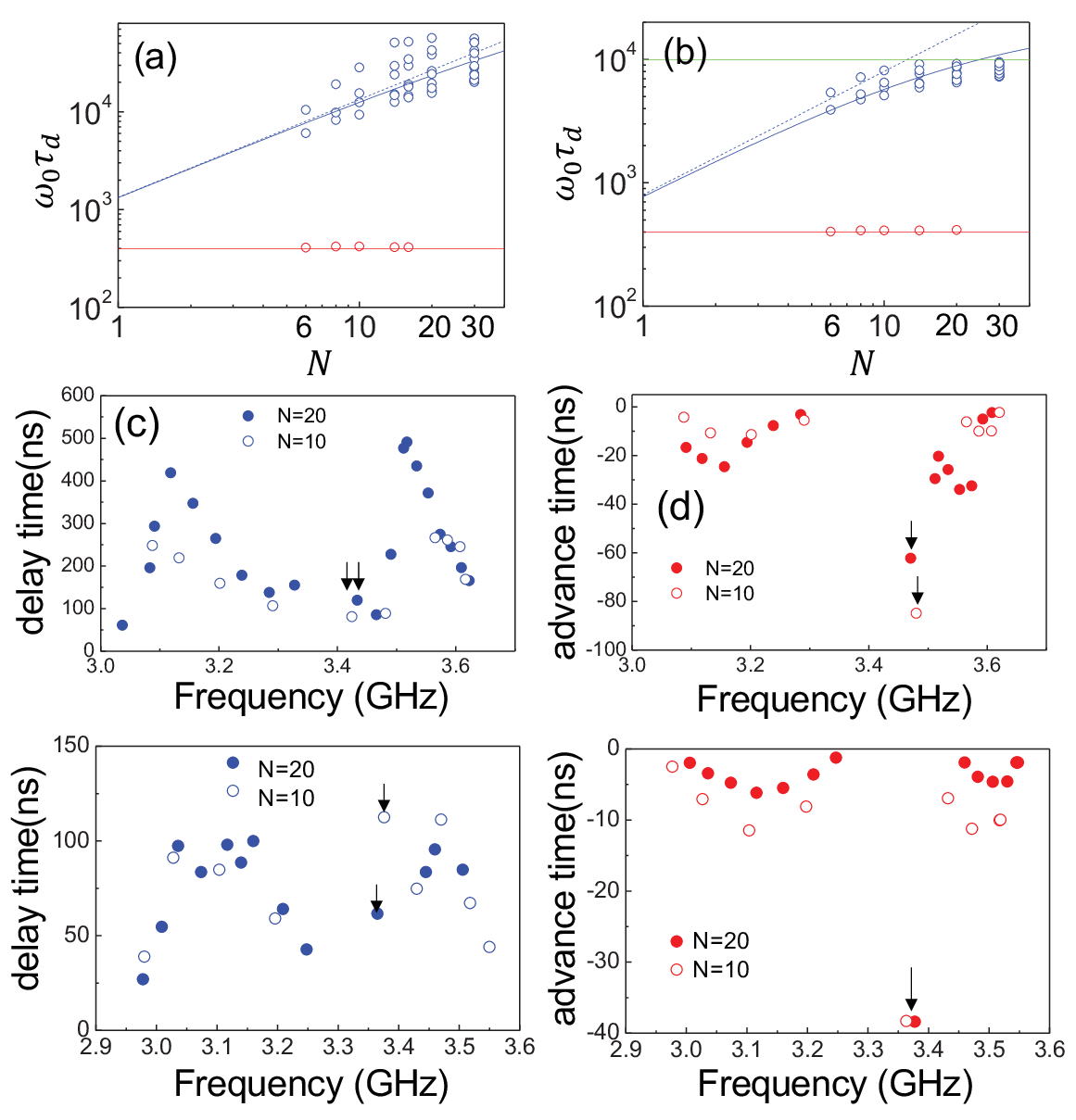}
\caption{(a) The delay time as a function of chain length $N$ for edge modes(red) and bulk mode(blue) determined by calculated transmission coefficient. The intrinsic loss and coupling rates are $\kappa_i/\omega_0=10^{-5}$ and $\kappa_{c0}/\omega_0=5\times10^{-3}$. (b) The delay time as a function of chain length $N$ with a higher intrinsic loss and coupling rates $\kappa_i/\omega_0=10^{-4}$ and $\kappa_{c0}/\omega_0=5\times10^{-3}$.
(c) The delay times deduced from the $S_{31}$ data obtained by finite-element simulations. (d) The advance time deduced from the $S_{21}$ data obtained by finite-element simulations. (e) The delay times deduced from the $S_{31}$ data from the $N=10$ and $N=20$ normal conductor samples. Vertical arrows indicate the results for edge modes. (f) The advance time deduced from the $S_{21}$ data from the $N=10$ and $N=20$ normal conductor samples. 
}
\end{figure}

Fig. 3(a)-(c) present the experimental results for the $N=20$ superconducting resonator chain. The plots show the reflection amplitude, transmission amplitude, and the advance/delay time at the eigenmodes, respectively. The measured band structure reveals a lower band ranging from 4.5 to 5.1 GHz, a higher band from 5.6 to 6.1 GHz, and edge mode in the band gap at 5.4 GHz. The pulse measurement indicates that the delay time for the edge mode is much smaller than those associated with the bulk modes. In this sample, the advance time is also long, with a magnitude comparable to the delay time. Furthermore, consistent with simulated results, both lead and delay times trend shorter as the frequency increases.

The delay time as a function of chain length $N$ for edge modes and bulk modes can be predicted by transmission coefficient calculations as presented in Fig. 4(a) and (b), respectively using $\kappa_i/\omega_0=10^{-5}$ and $10^{-4}$, and both  $\kappa_{c0}/\omega_0=5\times10^{-3}$. The red line shows the expected delay time for edge mode, $\tau_{d,m}=2/\kappa_{c0}$. The blue dashed curves indicates the linear dependence of $\tau_{d,m}$ and $N$, while the blue solid line includes the correction with a finite $\kappa_i$. The green line is the expected upper bound of delay time when the intrinsic loss is large. In Fig. 4 (c) and (d), we present the delay times obtained by finite-element simulations with the normal conductor strcutures with $N=10$ and 20. In simulations, the delay time of bulk mode transport increases with chain length, while the edge mode transport times remain similar for $N=10$ and $N=20$, as expected by theory for low-loss systems.

Experimentally, the length dependence of the transport times is investigated using the normal conducting samples, which have a larger intrinsic loss. In a high-loss regime where $\kappa_i > \sim \kappa_{c0}$, the delay time is dominated by the intrinsic loss $\tau_{d,m}\sim {2}/{\kappa_i},$ insensitive to the chain length $N$ for both bulk and edge modes. However, the scaling of the advance time can be deduced as $\tau_{a,m}\sim-2{\kappa_{c0}|\psi_1|^2}/{\kappa_i^2}$. For bulk modes, this advance time scales as $|\psi_1|^2\sim 1/N$, while for edge modes, $\tau_{a,m}$ is $N$-independent. As one can see in Fig. 4(e) and (f), the aforementioned expectations are verified except that the edge mode has an unexpected long delay time of 117.5 ns for $N=10$, that may be impacted by unwanted substrate modes. 

In conclusion, we have investigate the short transport time supported by the edge modes in low dimensional topological models, which is origniated from the large effective coupling between the edge mode and measurement leads. 
Regarding experimental demonstrations, we implement a chain of  interleaved split-ring resonators  and complementary split ring resonators with a proper setting in the orientation texture.
The existence of non-trivially topological edge states, which are zero modes, is not only demonstrated through the transmission spectra, but also verified through the observation of a short transport time.
The transport time revealed and verified in our experiments provides a crucial guideline for  utilizing photonic topological devices, which demonstrates the possibility to transport optical information with topological protection~\cite{T1, T2, T3, Mirhosseini18, T5}.

\section*{Acknowledgement}
We are grateful to the National Center for High-performance Computing for computer time and facilities. We have used the cryostat faciliity at Instrument Center, National Chung Hsing University. Fruitful discussions with M. Randeria, Dongning Zheng and H. Shimada are acknowledged. G.W. acknowledges the financial supports from National Natural Science Foundation of China (11604231); Natural Science Foundation of Jiangsu Province (BK20160303).This work is financially supported by the Ministry of Science and Technology, Taiwan under grant Nos.114-2112-M-005-011, 108-2923-M-007-001-MY3 and No. 109-2112-M-007-019-MY3), Office of Naval Research Global, US Army Research Office.

\bibstyle{prl}


\bibliography{srr_delay}

\begin{thebibliography}{30}%
\makeatletter
\providecommand \@ifxundefined [1]{%
 \@ifx{#1\undefined}
}%
\providecommand \@ifnum [1]{%
 \ifnum #1\expandafter \@firstoftwo
 \else \expandafter \@secondoftwo
 \fi
}%
\providecommand \@ifx [1]{%
 \ifx #1\expandafter \@firstoftwo
 \else \expandafter \@secondoftwo
 \fi
}%
\providecommand \natexlab [1]{#1}%
\providecommand \enquote  [1]{``#1''}%
\providecommand \bibnamefont  [1]{#1}%
\providecommand \bibfnamefont [1]{#1}%
\providecommand \citenamefont [1]{#1}%
\providecommand \href@noop [0]{\@secondoftwo}%
\providecommand \href [0]{\begingroup \@sanitize@url \@href}%
\providecommand \@href[1]{\@@startlink{#1}\@@href}%
\providecommand \@@href[1]{\endgroup#1\@@endlink}%
\providecommand \@sanitize@url [0]{\catcode `\\12\catcode `\$12\catcode
  `\&12\catcode `\#12\catcode `\^12\catcode `\_12\catcode `\%12\relax}%
\providecommand \@@startlink[1]{}%
\providecommand \@@endlink[0]{}%
\providecommand \url  [0]{\begingroup\@sanitize@url \@url }%
\providecommand \@url [1]{\endgroup\@href {#1}{\urlprefix }}%
\providecommand \urlprefix  [0]{URL }%
\providecommand \Eprint [0]{\href }%
\providecommand \doibase [0]{http://dx.doi.org/}%
\providecommand \selectlanguage [0]{\@gobble}%
\providecommand \bibinfo  [0]{\@secondoftwo}%
\providecommand \bibfield  [0]{\@secondoftwo}%
\providecommand \translation [1]{[#1]}%
\providecommand \BibitemOpen [0]{}%
\providecommand \bibitemStop [0]{}%
\providecommand \bibitemNoStop [0]{.\EOS\space}%
\providecommand \EOS [0]{\spacefactor3000\relax}%
\providecommand \BibitemShut  [1]{\csname bibitem#1\endcsname}%
\let\auto@bib@innerbib\@empty
\bibitem [{\citenamefont {Su}\ \emph {et~al.}(1980)\citenamefont {Su},
  \citenamefont {Schrieffer},\ and\ \citenamefont {Heeger}}]{SSH-1}%
  \BibitemOpen
  \bibfield  {author} {\bibinfo {author} {\bibfnamefont {W.~P.}\ \bibnamefont
  {Su}}, \bibinfo {author} {\bibfnamefont {J.~R.}\ \bibnamefont {Schrieffer}},
  \ and\ \bibinfo {author} {\bibfnamefont {A.~J.}\ \bibnamefont {Heeger}},\
  }\href@noop {} {\bibfield  {journal} {\bibinfo  {journal} {Physical Review
  B}\ }\textbf {\bibinfo {volume} {22}},\ \bibinfo {pages} {2099} (\bibinfo
  {year} {1980})}\BibitemShut {NoStop}%
\bibitem [{\citenamefont {Heeger}\ \emph {et~al.}(1988)\citenamefont {Heeger},
  \citenamefont {Kivelson}, \citenamefont {Schrieffer},\ and\ \citenamefont
  {Su}}]{SSH-2}%
  \BibitemOpen
  \bibfield  {author} {\bibinfo {author} {\bibfnamefont {A.~J.}\ \bibnamefont
  {Heeger}}, \bibinfo {author} {\bibfnamefont {S.}~\bibnamefont {Kivelson}},
  \bibinfo {author} {\bibfnamefont {J.~R.}\ \bibnamefont {Schrieffer}}, \ and\
  \bibinfo {author} {\bibfnamefont {W.~P.}\ \bibnamefont {Su}},\ }\href@noop {}
  {\bibfield  {journal} {\bibinfo  {journal} {Reviews of Modern Physics}\
  }\textbf {\bibinfo {volume} {60}},\ \bibinfo {pages} {781} (\bibinfo {year}
  {1988})}\BibitemShut {NoStop}%
\bibitem [{\citenamefont {Asb\'{o}th}\ \emph {et~al.}(2016)\citenamefont
  {Asb\'{o}th}, \citenamefont {Oroszl\'{a}ny},\ and\ \citenamefont
  {P\'{a}lyi}}]{book-SSH}%
  \BibitemOpen
  \bibfield  {author} {\bibinfo {author} {\bibfnamefont {J.~K.}\ \bibnamefont
  {Asb\'{o}th}}, \bibinfo {author} {\bibfnamefont {L.}~\bibnamefont
  {Oroszl\'{a}ny}}, \ and\ \bibinfo {author} {\bibfnamefont {A.}~\bibnamefont
  {P\'{a}lyi}},\ }\enquote {\bibinfo {title} {The su-schrieffer-heeger (ssh)
  model},}\ in\ \href@noop {} {\emph {\bibinfo {booktitle} {A Short Course on
  Topological Insulators: Band Structure and Edge States in One and Two
  Dimensions}}},\ \bibinfo {editor} {edited by\ \bibinfo {editor}
  {\bibfnamefont {J.~K.}\ \bibnamefont {Asb\'{o}th}}, \bibinfo {editor}
  {\bibfnamefont {L.}~\bibnamefont {Oroszl\'{a}ny}}, \ and\ \bibinfo {editor}
  {\bibfnamefont {A.}~\bibnamefont {P\'{a}lyi}}}\ (\bibinfo  {publisher}
  {Springer International Publishing},\ \bibinfo {address} {Cham},\ \bibinfo
  {year} {2016})\ pp.\ \bibinfo {pages} {1--22}\BibitemShut {NoStop}%
\bibitem [{\citenamefont {Malkova}\ \emph {et~al.}(2009)\citenamefont
  {Malkova}, \citenamefont {Hromada}, \citenamefont {Wang}, \citenamefont
  {Bryant},\ and\ \citenamefont {Chen}}]{OL-Chen}%
  \BibitemOpen
  \bibfield  {author} {\bibinfo {author} {\bibfnamefont {N.}~\bibnamefont
  {Malkova}}, \bibinfo {author} {\bibfnamefont {I.}~\bibnamefont {Hromada}},
  \bibinfo {author} {\bibfnamefont {X.}~\bibnamefont {Wang}}, \bibinfo {author}
  {\bibfnamefont {G.}~\bibnamefont {Bryant}}, \ and\ \bibinfo {author}
  {\bibfnamefont {Z.}~\bibnamefont {Chen}},\ }\href@noop {} {\bibfield
  {journal} {\bibinfo  {journal} {Optics Letters}\ }\textbf {\bibinfo {volume}
  {34}},\ \bibinfo {pages} {1633} (\bibinfo {year} {2009})}\BibitemShut
  {NoStop}%
\bibitem [{\citenamefont {Khanikaev}\ \emph {et~al.}(2013)\citenamefont
  {Khanikaev}, \citenamefont {Hossein~Mousavi}, \citenamefont {Tse},
  \citenamefont {Kargarian}, \citenamefont {MacDonald},\ and\ \citenamefont
  {Shvets}}]{TP-mat}%
  \BibitemOpen
  \bibfield  {author} {\bibinfo {author} {\bibfnamefont {A.~B.}\ \bibnamefont
  {Khanikaev}}, \bibinfo {author} {\bibfnamefont {S.}~\bibnamefont
  {Hossein~Mousavi}}, \bibinfo {author} {\bibfnamefont {W.-K.}\ \bibnamefont
  {Tse}}, \bibinfo {author} {\bibfnamefont {M.}~\bibnamefont {Kargarian}},
  \bibinfo {author} {\bibfnamefont {A.~H.}\ \bibnamefont {MacDonald}}, \ and\
  \bibinfo {author} {\bibfnamefont {G.}~\bibnamefont {Shvets}},\ }\href@noop {}
  {\bibfield  {journal} {\bibinfo  {journal} {Nat Mater}\ }\textbf {\bibinfo
  {volume} {12}},\ \bibinfo {pages} {233} (\bibinfo {year} {2013})}\BibitemShut
  {NoStop}%
\bibitem [{\citenamefont {Lu}\ \emph {et~al.}(2014)\citenamefont {Lu},
  \citenamefont {Joannopoulos},\ and\ \citenamefont {Soljacic}}]{TP-photon}%
  \BibitemOpen
  \bibfield  {author} {\bibinfo {author} {\bibfnamefont {L.}~\bibnamefont
  {Lu}}, \bibinfo {author} {\bibfnamefont {J.~D.}\ \bibnamefont
  {Joannopoulos}}, \ and\ \bibinfo {author} {\bibfnamefont {M.}~\bibnamefont
  {Soljacic}},\ }\href@noop {} {\bibfield  {journal} {\bibinfo  {journal} {Nat
  Photon}\ }\textbf {\bibinfo {volume} {8}},\ \bibinfo {pages} {821} (\bibinfo
  {year} {2014})}\BibitemShut {NoStop}%
\bibitem [{\citenamefont {Rechtsman}\ \emph {et~al.}(2013)\citenamefont
  {Rechtsman}, \citenamefont {Zeuner}, \citenamefont {Plotnik}, \citenamefont
  {Lumer}, \citenamefont {Podolsky}, \citenamefont {Dreisow}, \citenamefont
  {Nolte}, \citenamefont {Segev},\ and\ \citenamefont {Szameit}}]{TP-2D}%
  \BibitemOpen
  \bibfield  {author} {\bibinfo {author} {\bibfnamefont {M.~C.}\ \bibnamefont
  {Rechtsman}}, \bibinfo {author} {\bibfnamefont {J.~M.}\ \bibnamefont
  {Zeuner}}, \bibinfo {author} {\bibfnamefont {Y.}~\bibnamefont {Plotnik}},
  \bibinfo {author} {\bibfnamefont {Y.}~\bibnamefont {Lumer}}, \bibinfo
  {author} {\bibfnamefont {D.}~\bibnamefont {Podolsky}}, \bibinfo {author}
  {\bibfnamefont {F.}~\bibnamefont {Dreisow}}, \bibinfo {author} {\bibfnamefont
  {S.}~\bibnamefont {Nolte}}, \bibinfo {author} {\bibfnamefont
  {M.}~\bibnamefont {Segev}}, \ and\ \bibinfo {author} {\bibfnamefont
  {A.}~\bibnamefont {Szameit}},\ }\href@noop {} {\bibfield  {journal} {\bibinfo
   {journal} {Nature}\ }\textbf {\bibinfo {volume} {496}},\ \bibinfo {pages}
  {196} (\bibinfo {year} {2013})}\BibitemShut {NoStop}%
\bibitem [{\citenamefont {Cheng}\ \emph {et~al.}(2019)\citenamefont {Cheng},
  \citenamefont {Pan}, \citenamefont {Wang}, \citenamefont {Zhang},
  \citenamefont {Yu}, \citenamefont {Gover}, \citenamefont {Zhang},
  \citenamefont {Li}, \citenamefont {Zhou},\ and\ \citenamefont {Zhu}}]{Tao}%
  \BibitemOpen
  \bibfield  {author} {\bibinfo {author} {\bibfnamefont {Q.}~\bibnamefont
  {Cheng}}, \bibinfo {author} {\bibfnamefont {Y.}~\bibnamefont {Pan}}, \bibinfo
  {author} {\bibfnamefont {H.}~\bibnamefont {Wang}}, \bibinfo {author}
  {\bibfnamefont {C.}~\bibnamefont {Zhang}}, \bibinfo {author} {\bibfnamefont
  {D.}~\bibnamefont {Yu}}, \bibinfo {author} {\bibfnamefont {A.}~\bibnamefont
  {Gover}}, \bibinfo {author} {\bibfnamefont {H.}~\bibnamefont {Zhang}},
  \bibinfo {author} {\bibfnamefont {T.}~\bibnamefont {Li}}, \bibinfo {author}
  {\bibfnamefont {L.}~\bibnamefont {Zhou}}, \ and\ \bibinfo {author}
  {\bibfnamefont {S.}~\bibnamefont {Zhu}},\ }\href@noop {} {\bibfield
  {journal} {\bibinfo  {journal} {Physical Review Letters}\ }\textbf {\bibinfo
  {volume} {122}},\ \bibinfo {pages} {173901} (\bibinfo {year}
  {2019})}\BibitemShut {NoStop}%
\bibitem [{\citenamefont {Pilozzi}\ and\ \citenamefont {Conti}(2016)}]{laser1}%
  \BibitemOpen
  \bibfield  {author} {\bibinfo {author} {\bibfnamefont {L.}~\bibnamefont
  {Pilozzi}}\ and\ \bibinfo {author} {\bibfnamefont {C.}~\bibnamefont
  {Conti}},\ }\href@noop {} {\bibfield  {journal} {\bibinfo  {journal}
  {Physical Review B}\ }\textbf {\bibinfo {volume} {93}},\ \bibinfo {pages}
  {195317} (\bibinfo {year} {2016})}\BibitemShut {NoStop}%
\bibitem [{\citenamefont {Parto}\ \emph {et~al.}(2018)\citenamefont {Parto},
  \citenamefont {Wittek}, \citenamefont {Hodaei}, \citenamefont {Harari},
  \citenamefont {Bandres}, \citenamefont {Ren}, \citenamefont {Rechtsman},
  \citenamefont {Segev}, \citenamefont {Christodoulides},\ and\ \citenamefont
  {Khajavikhan}}]{laser}%
  \BibitemOpen
  \bibfield  {author} {\bibinfo {author} {\bibfnamefont {M.}~\bibnamefont
  {Parto}}, \bibinfo {author} {\bibfnamefont {S.}~\bibnamefont {Wittek}},
  \bibinfo {author} {\bibfnamefont {H.}~\bibnamefont {Hodaei}}, \bibinfo
  {author} {\bibfnamefont {G.}~\bibnamefont {Harari}}, \bibinfo {author}
  {\bibfnamefont {M.~A.}\ \bibnamefont {Bandres}}, \bibinfo {author}
  {\bibfnamefont {J.}~\bibnamefont {Ren}}, \bibinfo {author} {\bibfnamefont
  {M.~C.}\ \bibnamefont {Rechtsman}}, \bibinfo {author} {\bibfnamefont
  {M.}~\bibnamefont {Segev}}, \bibinfo {author} {\bibfnamefont {D.~N.}\
  \bibnamefont {Christodoulides}}, \ and\ \bibinfo {author} {\bibfnamefont
  {M.}~\bibnamefont {Khajavikhan}},\ }\href@noop {} {\bibfield  {journal}
  {\bibinfo  {journal} {Physical Review Letters}\ }\textbf {\bibinfo {volume}
  {120}},\ \bibinfo {pages} {113901} (\bibinfo {year} {2018})}\BibitemShut
  {NoStop}%
\bibitem [{\citenamefont {Marcucci}\ \emph {et~al.}(2019)\citenamefont
  {Marcucci}, \citenamefont {Pierangeli}, \citenamefont {Agranat},
  \citenamefont {Lee}, \citenamefont {DelRe},\ and\ \citenamefont
  {Conti}}]{control}%
  \BibitemOpen
  \bibfield  {author} {\bibinfo {author} {\bibfnamefont {G.}~\bibnamefont
  {Marcucci}}, \bibinfo {author} {\bibfnamefont {D.}~\bibnamefont
  {Pierangeli}}, \bibinfo {author} {\bibfnamefont {A.~J.}\ \bibnamefont
  {Agranat}}, \bibinfo {author} {\bibfnamefont {R.-K.}\ \bibnamefont {Lee}},
  \bibinfo {author} {\bibfnamefont {E.}~\bibnamefont {DelRe}}, \ and\ \bibinfo
  {author} {\bibfnamefont {C.}~\bibnamefont {Conti}},\ }\href@noop {}
  {\bibfield  {journal} {\bibinfo  {journal} {Nature Communications}\ }\textbf
  {\bibinfo {volume} {10}},\ \bibinfo {pages} {5090} (\bibinfo {year}
  {2019})}\BibitemShut {NoStop}%
\bibitem [{\citenamefont {Xia}\ \emph {et~al.}(2020)\citenamefont {Xia},
  \citenamefont {Juki\'{c}}, \citenamefont {Wang}, \citenamefont {Smirnova},
  \citenamefont {Smirnov}, \citenamefont {Tang}, \citenamefont {Song},
  \citenamefont {Szameit}, \citenamefont {Leykam}, \citenamefont {Xu},
  \citenamefont {Chen},\ and\ \citenamefont {Buljan}}]{nlin-SSH}%
  \BibitemOpen
  \bibfield  {author} {\bibinfo {author} {\bibfnamefont {S.}~\bibnamefont
  {Xia}}, \bibinfo {author} {\bibfnamefont {D.}~\bibnamefont {Juki\'{c}}},
  \bibinfo {author} {\bibfnamefont {N.}~\bibnamefont {Wang}}, \bibinfo {author}
  {\bibfnamefont {D.}~\bibnamefont {Smirnova}}, \bibinfo {author}
  {\bibfnamefont {L.}~\bibnamefont {Smirnov}}, \bibinfo {author} {\bibfnamefont
  {L.}~\bibnamefont {Tang}}, \bibinfo {author} {\bibfnamefont {D.}~\bibnamefont
  {Song}}, \bibinfo {author} {\bibfnamefont {A.}~\bibnamefont {Szameit}},
  \bibinfo {author} {\bibfnamefont {D.}~\bibnamefont {Leykam}}, \bibinfo
  {author} {\bibfnamefont {J.}~\bibnamefont {Xu}}, \bibinfo {author}
  {\bibfnamefont {Z.}~\bibnamefont {Chen}}, \ and\ \bibinfo {author}
  {\bibfnamefont {H.}~\bibnamefont {Buljan}},\ }\href@noop {} {\bibfield
  {journal} {\bibinfo  {journal} {Light: Science \& Applications}\ }\textbf
  {\bibinfo {volume} {9}},\ \bibinfo {pages} {147} (\bibinfo {year}
  {2020})}\BibitemShut {NoStop}%
\bibitem [{\citenamefont {Dai}\ \emph {et~al.}(2022)\citenamefont {Dai},
  \citenamefont {Ao}, \citenamefont {Bao}, \citenamefont {Mao}, \citenamefont
  {Chi}, \citenamefont {Fu}, \citenamefont {You}, \citenamefont {Chen},
  \citenamefont {Zhai}, \citenamefont {Tang}, \citenamefont {Yang},
  \citenamefont {Li}, \citenamefont {Yuan}, \citenamefont {Gao}, \citenamefont
  {Lin}, \citenamefont {Thompson}, \citenamefont {O’Brien}, \citenamefont
  {Li}, \citenamefont {Hu}, \citenamefont {Gong},\ and\ \citenamefont
  {Wang}}]{tp-emitter}%
  \BibitemOpen
  \bibfield  {author} {\bibinfo {author} {\bibfnamefont {T.}~\bibnamefont
  {Dai}}, \bibinfo {author} {\bibfnamefont {Y.}~\bibnamefont {Ao}}, \bibinfo
  {author} {\bibfnamefont {J.}~\bibnamefont {Bao}}, \bibinfo {author}
  {\bibfnamefont {J.}~\bibnamefont {Mao}}, \bibinfo {author} {\bibfnamefont
  {Y.}~\bibnamefont {Chi}}, \bibinfo {author} {\bibfnamefont {Z.}~\bibnamefont
  {Fu}}, \bibinfo {author} {\bibfnamefont {Y.}~\bibnamefont {You}}, \bibinfo
  {author} {\bibfnamefont {X.}~\bibnamefont {Chen}}, \bibinfo {author}
  {\bibfnamefont {C.}~\bibnamefont {Zhai}}, \bibinfo {author} {\bibfnamefont
  {B.}~\bibnamefont {Tang}}, \bibinfo {author} {\bibfnamefont {Y.}~\bibnamefont
  {Yang}}, \bibinfo {author} {\bibfnamefont {Z.}~\bibnamefont {Li}}, \bibinfo
  {author} {\bibfnamefont {L.}~\bibnamefont {Yuan}}, \bibinfo {author}
  {\bibfnamefont {F.}~\bibnamefont {Gao}}, \bibinfo {author} {\bibfnamefont
  {X.}~\bibnamefont {Lin}}, \bibinfo {author} {\bibfnamefont {M.~G.}\
  \bibnamefont {Thompson}}, \bibinfo {author} {\bibfnamefont {J.~L.}\
  \bibnamefont {O’Brien}}, \bibinfo {author} {\bibfnamefont {Y.}~\bibnamefont
  {Li}}, \bibinfo {author} {\bibfnamefont {X.}~\bibnamefont {Hu}}, \bibinfo
  {author} {\bibfnamefont {Q.}~\bibnamefont {Gong}}, \ and\ \bibinfo {author}
  {\bibfnamefont {J.}~\bibnamefont {Wang}},\ }\href@noop {} {\bibfield
  {journal} {\bibinfo  {journal} {Nature Photonics}\ }\textbf {\bibinfo
  {volume} {16}},\ \bibinfo {pages} {248} (\bibinfo {year} {2022})}\BibitemShut
  {NoStop}%
\bibitem [{\citenamefont {Falcone}\ \emph {et~al.}(2004)\citenamefont
  {Falcone}, \citenamefont {Lopetegi}, \citenamefont {Laso}, \citenamefont
  {Baena}, \citenamefont {Bonache}, \citenamefont {Beruete}, \citenamefont
  {Marques}, \citenamefont {Martin},\ and\ \citenamefont
  {Sorolla}}]{Falcone04}%
  \BibitemOpen
  \bibfield  {author} {\bibinfo {author} {\bibfnamefont {F.}~\bibnamefont
  {Falcone}}, \bibinfo {author} {\bibfnamefont {T.}~\bibnamefont {Lopetegi}},
  \bibinfo {author} {\bibfnamefont {M.~A.}\ \bibnamefont {Laso}}, \bibinfo
  {author} {\bibfnamefont {J.~D.}\ \bibnamefont {Baena}}, \bibinfo {author}
  {\bibfnamefont {J.}~\bibnamefont {Bonache}}, \bibinfo {author} {\bibfnamefont
  {M.}~\bibnamefont {Beruete}}, \bibinfo {author} {\bibfnamefont
  {R.}~\bibnamefont {Marques}}, \bibinfo {author} {\bibfnamefont
  {F.}~\bibnamefont {Martin}}, \ and\ \bibinfo {author} {\bibfnamefont
  {M.}~\bibnamefont {Sorolla}},\ }\href@noop {} {\bibfield  {journal} {\bibinfo
   {journal} {Phys Rev Lett}\ }\textbf {\bibinfo {volume} {93}},\ \bibinfo
  {pages} {197401} (\bibinfo {year} {2004})}\BibitemShut {NoStop}%
\bibitem [{\citenamefont {Shelby}\ \emph {et~al.}(2001)\citenamefont {Shelby},
  \citenamefont {Smith},\ and\ \citenamefont {Schultz}}]{Shelby01}%
  \BibitemOpen
  \bibfield  {author} {\bibinfo {author} {\bibfnamefont {R.}~\bibnamefont
  {Shelby}}, \bibinfo {author} {\bibfnamefont {D.}~\bibnamefont {Smith}}, \
  and\ \bibinfo {author} {\bibfnamefont {S.}~\bibnamefont {Schultz}},\
  }\href@noop {} {\bibfield  {journal} {\bibinfo  {journal} {Science}\ }\textbf
  {\bibinfo {volume} {292}},\ \bibinfo {pages} {77} (\bibinfo {year}
  {2001})}\BibitemShut {NoStop}%
\bibitem [{\citenamefont {Linden}\ \emph {et~al.}(2004)\citenamefont {Linden},
  \citenamefont {Enkrich}, \citenamefont {Wegener}, \citenamefont {Zhou},
  \citenamefont {Koschny},\ and\ \citenamefont {Soukoulis}}]{Linden04}%
  \BibitemOpen
  \bibfield  {author} {\bibinfo {author} {\bibfnamefont {S.}~\bibnamefont
  {Linden}}, \bibinfo {author} {\bibfnamefont {C.}~\bibnamefont {Enkrich}},
  \bibinfo {author} {\bibfnamefont {M.}~\bibnamefont {Wegener}}, \bibinfo
  {author} {\bibfnamefont {J.}~\bibnamefont {Zhou}}, \bibinfo {author}
  {\bibfnamefont {T.}~\bibnamefont {Koschny}}, \ and\ \bibinfo {author}
  {\bibfnamefont {C.~M.}\ \bibnamefont {Soukoulis}},\ }\href@noop {} {\bibfield
   {journal} {\bibinfo  {journal} {Science}\ }\textbf {\bibinfo {volume}
  {306}},\ \bibinfo {pages} {1351} (\bibinfo {year} {2004})}\BibitemShut
  {NoStop}%
\bibitem [{\citenamefont {Efremidis}(2021)}]{Efremidis}%
  \BibitemOpen
  \bibfield  {author} {\bibinfo {author} {\bibfnamefont {N.~K.}\ \bibnamefont
  {Efremidis}},\ }\href@noop {} {\bibfield  {journal} {\bibinfo  {journal}
  {Physical Review A}\ }\textbf {\bibinfo {volume} {104}},\ \bibinfo {pages}
  {053531} (\bibinfo {year} {2021})}\BibitemShut {NoStop}%
\bibitem [{\citenamefont {Kockum}\ \emph {et~al.}(2013)\citenamefont {Kockum},
  \citenamefont {Sandberg}, \citenamefont {Vissers}, \citenamefont {Gao},
  \citenamefont {Johansson},\ and\ \citenamefont {Pappas}}]{Kockum13}%
  \BibitemOpen
  \bibfield  {author} {\bibinfo {author} {\bibfnamefont {A.~F.}\ \bibnamefont
  {Kockum}}, \bibinfo {author} {\bibfnamefont {M.}~\bibnamefont {Sandberg}},
  \bibinfo {author} {\bibfnamefont {M.~R.}\ \bibnamefont {Vissers}}, \bibinfo
  {author} {\bibfnamefont {J.}~\bibnamefont {Gao}}, \bibinfo {author}
  {\bibfnamefont {G.}~\bibnamefont {Johansson}}, \ and\ \bibinfo {author}
  {\bibfnamefont {D.~P.}\ \bibnamefont {Pappas}},\ }\href@noop {} {\bibfield
  {journal} {\bibinfo  {journal} {Journal of Physics B: Atomic, Molecular and
  Optical Physics}\ }\textbf {\bibinfo {volume} {46}},\ \bibinfo {pages}
  {224014} (\bibinfo {year} {2013})}\BibitemShut {NoStop}%
\bibitem [{\citenamefont {Chien}\ \emph {et~al.}(2019)\citenamefont {Chien},
  \citenamefont {Hsieh}, \citenamefont {Chen}, \citenamefont {Dubyna},
  \citenamefont {Wu},\ and\ \citenamefont {Kuo}}]{Kuo19}%
  \BibitemOpen
  \bibfield  {author} {\bibinfo {author} {\bibfnamefont {W.-C.}\ \bibnamefont
  {Chien}}, \bibinfo {author} {\bibfnamefont {Y.-L.}\ \bibnamefont {Hsieh}},
  \bibinfo {author} {\bibfnamefont {C.-H.}\ \bibnamefont {Chen}}, \bibinfo
  {author} {\bibfnamefont {D.}~\bibnamefont {Dubyna}}, \bibinfo {author}
  {\bibfnamefont {C.-S.}\ \bibnamefont {Wu}}, \ and\ \bibinfo {author}
  {\bibfnamefont {W.}~\bibnamefont {Kuo}},\ }\href@noop {} {\bibfield
  {journal} {\bibinfo  {journal} {Opt. Express}\ }\textbf {\bibinfo {volume}
  {27}},\ \bibinfo {pages} {36088} (\bibinfo {year} {2019})}\BibitemShut
  {NoStop}%
\bibitem [{\citenamefont {Chang}\ \emph {et~al.}(2024)\citenamefont {Chang},
  \citenamefont {Abdelghany}, \citenamefont {Peng}, \citenamefont {Wu},\ and\
  \citenamefont {Kuo}}]{Chang24}%
  \BibitemOpen
  \bibfield  {author} {\bibinfo {author} {\bibfnamefont {Y.-H.}\ \bibnamefont
  {Chang}}, \bibinfo {author} {\bibfnamefont {R.~A.}\ \bibnamefont
  {Abdelghany}}, \bibinfo {author} {\bibfnamefont {W.~L.}\ \bibnamefont
  {Peng}}, \bibinfo {author} {\bibfnamefont {C.-S.}\ \bibnamefont {Wu}}, \ and\
  \bibinfo {author} {\bibfnamefont {W.}~\bibnamefont {Kuo}},\ }\href@noop {}
  {\bibfield  {journal} {\bibinfo  {journal} {Optics Express}\ }\textbf
  {\bibinfo {volume} {32}},\ \bibinfo {pages} {30955} (\bibinfo {year}
  {2024})}\BibitemShut {NoStop}%
\bibitem [{\citenamefont {Sersic}\ \emph {et~al.}(2009)\citenamefont {Sersic},
  \citenamefont {Frimmer}, \citenamefont {Verhagen},\ and\ \citenamefont
  {Koenderink}}]{Ser09}%
  \BibitemOpen
  \bibfield  {author} {\bibinfo {author} {\bibfnamefont {I.}~\bibnamefont
  {Sersic}}, \bibinfo {author} {\bibfnamefont {M.}~\bibnamefont {Frimmer}},
  \bibinfo {author} {\bibfnamefont {E.}~\bibnamefont {Verhagen}}, \ and\
  \bibinfo {author} {\bibfnamefont {A.~F.}\ \bibnamefont {Koenderink}},\
  }\href@noop {} {\bibfield  {journal} {\bibinfo  {journal} {Physical Review
  Letters}\ }\textbf {\bibinfo {volume} {103}},\ \bibinfo {pages} {213902}
  (\bibinfo {year} {2009})}\BibitemShut {NoStop}%
\bibitem [{\citenamefont {Seetharaman}\ \emph {et~al.}(2017)\citenamefont
  {Seetharaman}, \citenamefont {King}, \citenamefont {Hooper},\ and\
  \citenamefont {Barnes}}]{See17}%
  \BibitemOpen
  \bibfield  {author} {\bibinfo {author} {\bibfnamefont {S.~S.}\ \bibnamefont
  {Seetharaman}}, \bibinfo {author} {\bibfnamefont {C.~G.}\ \bibnamefont
  {King}}, \bibinfo {author} {\bibfnamefont {I.~R.}\ \bibnamefont {Hooper}}, \
  and\ \bibinfo {author} {\bibfnamefont {W.~L.}\ \bibnamefont {Barnes}},\
  }\href@noop {} {\bibfield  {journal} {\bibinfo  {journal} {Physical Review
  B}\ }\textbf {\bibinfo {volume} {96}},\ \bibinfo {pages} {085426} (\bibinfo
  {year} {2017})}\BibitemShut {NoStop}%
\bibitem [{\citenamefont {Jiang}\ \emph {et~al.}(2018)\citenamefont {Jiang},
  \citenamefont {Guo}, \citenamefont {Ding}, \citenamefont {Sun}, \citenamefont
  {Li}, \citenamefont {Jiang},\ and\ \citenamefont {Chen}}]{Jiang}%
  \BibitemOpen
  \bibfield  {author} {\bibinfo {author} {\bibfnamefont {J.}~\bibnamefont
  {Jiang}}, \bibinfo {author} {\bibfnamefont {Z.}~\bibnamefont {Guo}}, \bibinfo
  {author} {\bibfnamefont {Y.}~\bibnamefont {Ding}}, \bibinfo {author}
  {\bibfnamefont {Y.}~\bibnamefont {Sun}}, \bibinfo {author} {\bibfnamefont
  {Y.}~\bibnamefont {Li}}, \bibinfo {author} {\bibfnamefont {H.}~\bibnamefont
  {Jiang}}, \ and\ \bibinfo {author} {\bibfnamefont {H.}~\bibnamefont {Chen}},\
  }\href@noop {} {\bibfield  {journal} {\bibinfo  {journal} {Optics Express}\
  }\textbf {\bibinfo {volume} {26}},\ \bibinfo {pages} {12891} (\bibinfo {year}
  {2018})}\BibitemShut {NoStop}%
\bibitem [{\citenamefont {Lin}\ \emph {et~al.}(2017)\citenamefont {Lin},
  \citenamefont {Chang}, \citenamefont {Chien},\ and\ \citenamefont
  {Kuo}}]{Lin17}%
  \BibitemOpen
  \bibfield  {author} {\bibinfo {author} {\bibfnamefont {Y.-J.}\ \bibnamefont
  {Lin}}, \bibinfo {author} {\bibfnamefont {Y.-H.}\ \bibnamefont {Chang}},
  \bibinfo {author} {\bibfnamefont {W.-C.}\ \bibnamefont {Chien}}, \ and\
  \bibinfo {author} {\bibfnamefont {W.}~\bibnamefont {Kuo}},\ }\href@noop {}
  {\bibfield  {journal} {\bibinfo  {journal} {Optics Express}\ }\textbf
  {\bibinfo {volume} {25}},\ \bibinfo {pages} {30395} (\bibinfo {year}
  {2017})}\BibitemShut {NoStop}%
\bibitem [{\citenamefont {Chang}\ \emph {et~al.}(2023)\citenamefont {Chang},
  \citenamefont {Silalahi}, \citenamefont {Yang}, \citenamefont {Wen},\ and\
  \citenamefont {Kuo}}]{Chang23}%
  \BibitemOpen
  \bibfield  {author} {\bibinfo {author} {\bibfnamefont {Y.-H.}\ \bibnamefont
  {Chang}}, \bibinfo {author} {\bibfnamefont {V.~C.}\ \bibnamefont {Silalahi}},
  \bibinfo {author} {\bibfnamefont {Y.-T.}\ \bibnamefont {Yang}}, \bibinfo
  {author} {\bibfnamefont {Y.-S.}\ \bibnamefont {Wen}}, \ and\ \bibinfo
  {author} {\bibfnamefont {W.}~\bibnamefont {Kuo}},\ }\href@noop {} {\bibfield
  {journal} {\bibinfo  {journal} {Optics Express}\ }\textbf {\bibinfo {volume}
  {31}},\ \bibinfo {pages} {24492} (\bibinfo {year} {2023})}\BibitemShut
  {NoStop}%
\bibitem [{\citenamefont {Mittal}\ \emph {et~al.}(2014)\citenamefont {Mittal},
  \citenamefont {Fan}, \citenamefont {Faez}, \citenamefont {Migdall},
  \citenamefont {Taylor},\ and\ \citenamefont {Hafezi}}]{T1}%
  \BibitemOpen
  \bibfield  {author} {\bibinfo {author} {\bibfnamefont {S.}~\bibnamefont
  {Mittal}}, \bibinfo {author} {\bibfnamefont {J.}~\bibnamefont {Fan}},
  \bibinfo {author} {\bibfnamefont {S.}~\bibnamefont {Faez}}, \bibinfo {author}
  {\bibfnamefont {A.}~\bibnamefont {Migdall}}, \bibinfo {author} {\bibfnamefont
  {J.~M.}\ \bibnamefont {Taylor}}, \ and\ \bibinfo {author} {\bibfnamefont
  {M.}~\bibnamefont {Hafezi}},\ }\href@noop {} {\bibfield  {journal} {\bibinfo
  {journal} {Physical Review Letters}\ }\textbf {\bibinfo {volume} {113}},\
  \bibinfo {pages} {087403} (\bibinfo {year} {2014})}\BibitemShut {NoStop}%
\bibitem [{\citenamefont {Blanco-Redondo}\ \emph {et~al.}(2016)\citenamefont
  {Blanco-Redondo}, \citenamefont {Andonegui}, \citenamefont {Collins},
  \citenamefont {Harari}, \citenamefont {Lumer}, \citenamefont {Rechtsman},
  \citenamefont {Eggleton},\ and\ \citenamefont {Segev}}]{T2}%
  \BibitemOpen
  \bibfield  {author} {\bibinfo {author} {\bibfnamefont {A.}~\bibnamefont
  {Blanco-Redondo}}, \bibinfo {author} {\bibfnamefont {I.}~\bibnamefont
  {Andonegui}}, \bibinfo {author} {\bibfnamefont {M.~J.}\ \bibnamefont
  {Collins}}, \bibinfo {author} {\bibfnamefont {G.}~\bibnamefont {Harari}},
  \bibinfo {author} {\bibfnamefont {Y.}~\bibnamefont {Lumer}}, \bibinfo
  {author} {\bibfnamefont {M.~C.}\ \bibnamefont {Rechtsman}}, \bibinfo {author}
  {\bibfnamefont {B.~J.}\ \bibnamefont {Eggleton}}, \ and\ \bibinfo {author}
  {\bibfnamefont {M.}~\bibnamefont {Segev}},\ }\href@noop {} {\bibfield
  {journal} {\bibinfo  {journal} {Physical Review Letters}\ }\textbf {\bibinfo
  {volume} {116}},\ \bibinfo {pages} {163901} (\bibinfo {year}
  {2016})}\BibitemShut {NoStop}%
\bibitem [{\citenamefont {Weimann}\ \emph {et~al.}(2017)\citenamefont
  {Weimann}, \citenamefont {Kremer}, \citenamefont {Plotnik}, \citenamefont
  {Lumer}, \citenamefont {Nolte}, \citenamefont {Makris}, \citenamefont
  {Segev}, \citenamefont {Rechtsman},\ and\ \citenamefont {Szameit}}]{T3}%
  \BibitemOpen
  \bibfield  {author} {\bibinfo {author} {\bibfnamefont {S.}~\bibnamefont
  {Weimann}}, \bibinfo {author} {\bibfnamefont {M.}~\bibnamefont {Kremer}},
  \bibinfo {author} {\bibfnamefont {Y.}~\bibnamefont {Plotnik}}, \bibinfo
  {author} {\bibfnamefont {Y.}~\bibnamefont {Lumer}}, \bibinfo {author}
  {\bibfnamefont {S.}~\bibnamefont {Nolte}}, \bibinfo {author} {\bibfnamefont
  {K.~G.}\ \bibnamefont {Makris}}, \bibinfo {author} {\bibfnamefont
  {M.}~\bibnamefont {Segev}}, \bibinfo {author} {\bibfnamefont {M.~C.}\
  \bibnamefont {Rechtsman}}, \ and\ \bibinfo {author} {\bibfnamefont
  {A.}~\bibnamefont {Szameit}},\ }\href@noop {} {\bibfield  {journal} {\bibinfo
   {journal} {Nature Materials}\ }\textbf {\bibinfo {volume} {16}},\ \bibinfo
  {pages} {433} (\bibinfo {year} {2017})}\BibitemShut {NoStop}%
\bibitem [{\citenamefont {Mirhosseini}\ \emph {et~al.}(2018)\citenamefont
  {Mirhosseini}, \citenamefont {Kim}, \citenamefont {Ferreira}, \citenamefont
  {Kalaee}, \citenamefont {Sipahigil}, \citenamefont {Keller},\ and\
  \citenamefont {Painter}}]{Mirhosseini18}%
  \BibitemOpen
  \bibfield  {author} {\bibinfo {author} {\bibfnamefont {M.}~\bibnamefont
  {Mirhosseini}}, \bibinfo {author} {\bibfnamefont {E.}~\bibnamefont {Kim}},
  \bibinfo {author} {\bibfnamefont {V.~S.}\ \bibnamefont {Ferreira}}, \bibinfo
  {author} {\bibfnamefont {M.}~\bibnamefont {Kalaee}}, \bibinfo {author}
  {\bibfnamefont {A.}~\bibnamefont {Sipahigil}}, \bibinfo {author}
  {\bibfnamefont {A.~J.}\ \bibnamefont {Keller}}, \ and\ \bibinfo {author}
  {\bibfnamefont {O.}~\bibnamefont {Painter}},\ }\href@noop {} {\bibfield
  {journal} {\bibinfo  {journal} {Nature Communications}\ }\textbf {\bibinfo
  {volume} {9}},\ \bibinfo {pages} {3706} (\bibinfo {year} {2018})}\BibitemShut
  {NoStop}%
\bibitem [{\citenamefont {Arregui}\ \emph {et~al.}(2021)\citenamefont
  {Arregui}, \citenamefont {Gomis-Bresco}, \citenamefont {Sotomayor-Torres},\
  and\ \citenamefont {Garcia}}]{T5}%
  \BibitemOpen
  \bibfield  {author} {\bibinfo {author} {\bibfnamefont {G.}~\bibnamefont
  {Arregui}}, \bibinfo {author} {\bibfnamefont {J.}~\bibnamefont
  {Gomis-Bresco}}, \bibinfo {author} {\bibfnamefont {C.~M.}\ \bibnamefont
  {Sotomayor-Torres}}, \ and\ \bibinfo {author} {\bibfnamefont {P.~D.}\
  \bibnamefont {Garcia}},\ }\href@noop {} {\bibfield  {journal} {\bibinfo
  {journal} {Physical Review Letters}\ }\textbf {\bibinfo {volume} {126}},\
  \bibinfo {pages} {027403} (\bibinfo {year} {2021})}\BibitemShut {NoStop}%
\end{thebibliography}%

\end{document}